c	             prb.tex
\documentstyle[prb,aps,preprint]{revtex} 
\begin{document}  
\draft

\title{A $^4$He shadow wavefunction with an inverse seventh power 
particle-particle correlation function}

\author{Orion Ciftja and Siu A. Chin}
\address{Department of Physics, Texas $A\&M$ University, College
        Station, Texas 77843 }
\author{Francesco Pederiva}
\address{Dipartimento di Fisica, Universit\'{a} di Trento, Povo,
         Trento, Italy }
\date{\today}   
\maketitle

\begin{abstract}
Many ground state studies of $^4$He using a shadow wave function with 
an inverse fifth power McMillan particle-particle correlation 
function have yielded radial distribution functions 
with misplaced peaks. It has been conjectured that this is due to 
the specific choice of the McMillan correlation function.
However, beyond the use of fully optimized two-particle correlation 
functions, there has been little study of simple alternatives
that can correct this defect.
In this work we show that the remedy is surprisingly simple.
When a shadow wavefunction with an inverse seventh power particle-particle 
correlation function is used to study $^4$He, it gives
a correctly peaked radial distribution function, lowers the energy
at all liquid and solid densities, and produces an excellent equation of
state.

\end{abstract}


The ground state properties of liquid and solid $^4$He
have been studied extensively over the years by a variety of 
many-body techniques, ranging from hypernetted-chain (HNC) 
theories~\cite{usmani,krotscheck} and variational Monte Carlo (VMC) 
methods~\cite{mcmillan,triplet} to ``exact" Monte Carlo methods (EMC) such
as Green's Function Monte Carlo\cite{ceperley,gfmc} (GFMC) 
and Diffusion Monte Carlo\cite{boronat94} (DMC).
The advance of EMC have seemingly oblivated
the need for the ``brutish" and bias-laiden method of VMC. 
However, the introduction of the shadow wavefunction by 
Vitiello {\it et al}\,\cite{vitiello}
has added new subtlety and refinement to this approach. It has been 
shown that shadow wavefunctions can describe both the liquid
and solid phase of $^4$He with excellent energy, while simultaneously
maintaining translational and Bose sysmmetry. 
Since any VMC calculation 
is an order of magnitude less computationally demanding than corresponding 
EMC calculations, the method of shadow wavefunctions remained economically
and conceptionally appealing. 

However, it has been noted for some time that shadow wavefunctions with
the McMillan inverse fifth power particle-particle correlation 
function do not give a correct radial distribution function for bulk
liquid $^4$He. All the peaks are misplaced as in the original McMillan
calculation. While there have been continued improvements on the form of
two-particle correlation, leading to fully optimized correlations 
expressible in terms of a basis state, there has been no 
explorations of simpler alternatives to cure this defect. In this work, 
we shown that a simple replacement of the inverse fifth power by that
of an inverse seventh power significantly improves simultaneously, the 
ground state energy, the equation of state and the 
radial distribution function. 

The first VMC calculation of the groundstate
properties of liquid and solid $^4$He was carried out by
McMillan~\cite{mcmillan} who employed a trial wave function
with an inverse fifth power of the particle separation as the 
two-body correlation function.
In this early study the potential between $^4$He atoms 
is taken to be the two-body Lennard-Jones (LJ) potential,
$v(r)=4 \epsilon [ (\sigma/r)^{12}-(\sigma/r)^{6} ]$, with the
DeBoer-Michels parameters, $\sigma=2.556 \AA$ and 
$\epsilon=10.22 K^o$. Since then, the two-body HFDHE2 potential 
of Aziz {\it et al}\,\cite{aziz79} has 
superceded the LJ potential as the potential of choice 
for the Helium studies.
More recent minor revisions of this potential\cite{aziz2}
have added, but have not greatly altered the quality 
of description of the interaction between Helium atoms 
at low pressure. In this work, we will continue to
use the HFDHE2 potential to facilitate comparison with existing
calculations in the literature.

The original shadow wavefunction of Vitiello {\it et al}\,\cite{vitiello}
can be written as:
\begin{equation}
\Psi( {\bf R}) =\Phi( {\bf R}) \int d  {\bf S} \ \Theta( {\bf R}, {\bf S})
              \  \Phi_S( {\bf S})   \ ,
\label{newswf}
\end{equation}
with a Gaussian particle-shadow correlation function 
\begin{equation}
\Theta({\bf R},{\bf S})=
    \exp \left[ -\sum_{i=1}^{N} C ( {\bf r}_{i}- {\bf s}_{i})^2 \right] \ ,
\label{Theta}
\end{equation}
%
where $ {\bf R} \equiv \{  {\bf r}_1, {\bf r}_2, \ldots , {\bf r}_N \}$ and
$ {\bf S} \equiv \{  {\bf s}_1, {\bf s}_2, \ldots , {\bf s}_N \}$ represent the
set of particle and shadow coordinates.
The particle-particle correlation function $\Phi( {\bf R})$ and the
shadow-shadow correlation function $\Phi_S( {\bf S})$ are both
of the pair-wise product form, 
$\Phi( {\bf R})=\exp\left[-\sum_{i>j}^{N} u(r_{ij}) \right]$ and
$\Phi_S( {\bf S})=\exp\left[-\sum_{i>j}^{N} u_S(s_{ij}) \right]$, where
$r_{ij}=| {\bf r}_i- {\bf r}_j|$ and $s_{ij}=| {\bf s}_i- {\bf s}_j|$.
Both $\Phi( {\bf R})$ and $\Phi_S( {\bf S})$ are
taken as McMillan form of inverse fifth powers (m=5) with
$u(r)=\frac{1}{2} (\frac{b}{r})^m$ and
$u_S(s)=(\frac{b^\prime}{s})^m$, where $b$ and $b^\prime$ are two
variational parameters.
The one-to-one coupling constant $C$ between particles and shadows 
is  treated as a variational parameter. We refer to this wavefunction as M+MS.

Optimizing the shadow-shadow 
pseudopotential~\cite{vitielloprb}
in the form of
$u_S(s)=(\frac{b^\prime}{s})^n$ 
with two variational parameters $b^\prime$ and $n$, produces no
significant improvements. A better choice was found  
following a suggestion by Reatto {\it et al}\,\cite{reatto} :
$u_S(s)= \tau \ v(\alpha s)$, where $v(r)$ is the Aziz HFDHE2 potential
and $\tau$ and $\alpha$ are variational parameters.
This scaled Aziz shadow correlation, introduced by 
MacFarland {\it et al}\,\cite{macfarland},
which uses a McMillan inverse fifth power
particle-particle psedopotential (m=5), will be denoted as M+AS.
The M+AS shadow wavefunction
substantially lowered the variational energies and improved the description 
of liquid and solid $^4$He at all densities. For example, at
the GFMC equilibrium density $\rho \sigma^3=0.365$ and at 
freezing density of $\rho \sigma^3=0.438$, the
M+AS energy is about 0.5 $K^o$ lower than the M+MS energy.

Both wavefunctions, however, produce a radial distribution function
at equilibrium density whose main peak is shifted outward by 
about 0.1 $\AA$ as compared with the experimental value. The same
misplacement is also observed at the GFMC freezing density. Such
a misplacement can be corrected by optimizing the two-body 
correlations through the method of basis state expansion~\cite{basisset}. 
(The peak height is still underestimated, however.)

In this work, we show that this crucial defect can be simply corrected
by a better choice of the inverse power (from 5 to 7) in McMillan's form of the
particle-particle correlation function.

Without reoptimizing the M+AS wavefunction's shadow parameters, 
but only varying the variational parameter $b$ with $m=7$,
we obtained lower energies
than those of M+AS at all liquid and solid densities, in an
amount ranging from $0.1$ to $0.3 K^o$.
This choice of the wavefunction in our work, 
referred to as M7+AS, allows us to improve the quality of the 
shadow wavefunction while retaining the same level of simplicity as before.

For a system of $N$ Helium atoms interacting via two-body forces
only, the Hamiltonian has the form

\begin{equation}
\hat{H}=-\frac{\hbar^2}{2m} \sum_{i=1}^{N} \nabla_{i}^{2}+
         \sum_{i>j}^{N} v(r_{ij}) \ .
\label{eqH}
\end{equation}
where $v(r_{ij})$ is the Aziz HFDHE2 potential. The expectation value 
of the Hamiltonian can be expressed as
\begin{eqnarray}
E=&&\frac{\int d {\bf R} \Psi( {\bf R}) \hat{H} \Psi( {\bf R})}{\int d {\bf R} 
   |\Psi( {\bf R})|^2}\\
 =&&\int d {\bf R} d {\bf S} d {\bf S^{\prime}} \
    p( {\bf R}, {\bf S}, {\bf S^{\prime}}) \
    E_L( {\bf R}, {\bf S}, {\bf S^{\prime}}) \ .
\label{energy2}
\end{eqnarray}
The local energy is written as
\begin{equation}
E_L( {\bf R}, {\bf S}, {\bf S^{\prime}})=
 \frac{\hat{H} \  \Phi( {\bf R}) \Theta( {\bf R}, {\bf S})}{
 \Phi( {\bf R}) \Theta( {\bf R}, {\bf S})}  \ ,
\label{elocal}
\end{equation}
and does not depend on $\Phi_S( {\bf S})$ since $\hat{H}$ acts only upon
the variables describing the system of real particles.

The probability $ p( {\bf R}, {\bf S}, {\bf S^{\prime}})$ is given by
\begin{eqnarray}
\lefteqn{ p( {\bf R}, {\bf S}, {\bf S^{\prime}})=} \nonumber \\
 & & \frac{\Phi( {\bf R})^2 \ \Theta( {\bf R}, {\bf S}) \ \Phi_S( {\bf S})
\ \Theta( {\bf R}, {\bf S^{\prime}}) \ \Phi_S( {\bf S^{\prime}})}{
\int d {\bf R} \ d {\bf S} \ d {\bf S^{\prime}} \
\Phi( {\bf R})^2 \ \Theta( {\bf R}, {\bf S}) \ \Phi_S( {\bf S})
\ \Theta( {\bf R}, {\bf S^{\prime}}) \ \Phi_S( {\bf S^{\prime}})} 
\label{prob}
\end{eqnarray}

To evaluate the expectation value of the Hamiltonian we use the Metropolis
Monte Carlo algorithm~\cite{metropolis} to sample the probability density
$p( {\bf R}, {\bf S}, {\bf S^{\prime}})$ from the $9N$ dimensional configuration
space of the particles and two sets of shadow coordinates.
In these computations, the Metropolis steps are subdivided in two
parts. In the first, one attempts to move real particle coordinates at random
inside cubical boxes of side length $\Delta$. In the second, analogous 
attempts to move shadow coordinates are made inside cubical boxes of
side length $\Delta^\prime$.
After we attempt to move all the shadow coordinates of set $\{  {\bf S} \}$,
the same is done for those in set $\{  {\bf S^{\prime}} \}$.
The parameters $\Delta$ and $\Delta^\prime$ were adjusted so that the
acceptance ratio for both particle and shadow moves was nearly $50\%$.

We compute the ground state variational energy,
the radial distribution function $g(r)$, and the static structure
factor $S(k)$.
These quantities are spherical averages and have been computed for both
the real particles and the shadow coordinates.
The radial distribution function is defined by
\begin{equation}
g(r)=\frac{1}{N \rho} \sum_{i \neq j}^{N} 
     \langle \delta(| {\bf r}_i- {\bf r}_j- {\bf r}|) \rangle  \ ,
\label{radial}
\end{equation}
where the angular brackets denote an average with respect to
$|\Psi( {\bf R})|^2$ and $\rho$ is the particle density.
The static structure factor is obtained from the average
\begin{equation}
S(k)=\frac{1}{N} \langle \rho_{- {\bf k}} \rho_{ {\bf k}} \rangle \ ,
\label{fluct}
\end{equation}
where $\rho_{ {\bf k}}$ is given by
$\rho_{ {\bf k}}= \sum_{j=1}^{N} \exp(-i  {\bf k}  {\bf r}_j)$.
By using this procedure $S(k)$ is computed for 
a discrete set of $ {\bf k}$ values where the smaller wave vector 
compatible with the periodic boundary condition of the 
system is $k=2 \pi/L$ ( $L$ is the side of the simulation box ).

%
%
%

All simulations presented in this work have been done with $N=108$
atoms of $^4$He in a cubic box with periodic boundary conditions.
To enforce periodicity, the two-body interaction potential $v(r)$ 
smoothly goes to zero at a cutoff distance, $r_c=L/2$, 
equal half the side of the simulation box. We actually use a
slightly modified two-body interaction potential 
$v^{\prime}(r) \equiv v(r)- \Delta v(r) $ according to the replacement

\begin{equation}
v^{\prime}(r)= \left \{
\begin{array}{ll}
v(r)+v(2 r_c-r)-2 v(r_c) ,     & \mbox {$ r \le r_{c} $} \\
0,                     & \mbox { $r>r_c$ }                         \ .
\end{array}
	                                 \right.
\label{vprime}
\end{equation}
A correction $ \Delta V=(\rho/2) \int d^3r g(r) \Delta v(r)$ was then
added to the computed potential energy, where the radial distribution
function $g(r)$ comes from the simulation and is taken equal to $1$ 
for $r>r_c$.
The shadow-shadow pseudopotential $u_S(s)$ was modified according to the
same prescription as $v(r)$, while the particle-particle
inverse power McMillan pseudopotential $u(r)$ and its first two derivatives
were slightly modified near the edge of the simulation box in order
to go smoothly to zero, by using a third degree polynomial fit
to the pseudopotential near the edge of the simulation box.

All calculations start from a perfect fcc crystal.
Our runs consisted of a total of about $5.5 \cdot 10^5$ passes during each
of which an attempt was made to move particles and shadows.
We allowed about $ 50 \cdot 10^3$ passes for equilibration followed
by about $5 \cdot 10^5$ passes which comprise the equilibrated random
walk.

In  Table~\ref{tabliquid} we show the energy per particle obtained
from the M7+AS shadow wavefunction after 
simulations with $N=108$ particles at several densities of
liquid $^4$He.
Also included in the table are several results from the literature,
GFMC refers to the results of Kalos {\it et al}\,\cite{gfmc}.

In Table~\ref{parliq} we show the values of the optimum variational
parameters $b$, $C$, $\tau$, and $\alpha$ for the M7+AS shadow
wavefunction at different densities $\rho$ in the liquid phase.

The energy per particle for the M7+AS shadow wavefunction after 
simulations with $N=108$ particles at some densities in the solid phase 
is shown in Table~\ref{tabsolid}. In the same table we show the VMC 
results obtained with the M+MS and M+AS shadow wavefunction, 
as well as the GFMC results.

Table~\ref{parsol} shows the values of the optimum variational
parameters $b$, $C$, $\tau$, and $\alpha$ for the M7+AS shadow
wavefunction at different densities $\rho$ in the solid phase.
%
%
%
We fit our equation of state in the liquid phase to a cubic
polynomial of the form
\begin{equation}
E(\rho)=E_0+B \left[ \frac{ \rho-\rho_0}{\rho_0} \right]^2+
           C \left[ \frac{ \rho-\rho_0}{\rho_0} \right]^3 \ ,
\label{pol}
\end{equation}
where $\rho_0$ is the equilibrium density.
A similar function has been used to fit the experimental equation
of state~\cite{roach} and we use it to analyze our M7+AS results.
The values of the parameters in the fit and their errors are shown in
Table~\ref{fitliq}, together with other results from the literature.
One notes that the values of the coefficients $B$, $C$, and $\rho_0 \sigma^3$
for the M7+AS shadow wavefunction are in good agreement with both
GFMC and experimental results~\cite{roach}.
In the solid phase we used the same parametrization as reported
for the M+AS case. We fited the energy to a cubic polynomial of the form
\begin{equation}
E(\rho)=E_0+B \left[ \frac{ \rho-\rho_S}{\rho_S} \right]^2+
           C \left[ \frac{ \rho-\rho_S}{\rho_S} \right]^3 \ ,
\label{polsol}
\end{equation}
where the specific density $\rho_S \sigma^3$=0.4486 is taken from the
GFMC calculation.
The values of the parameters to this fit are shown in Table~\ref{fitsol},
together with other results from the literature.
Again the M7+AS shadow wavefunction shows a good agreement with GFMC
with the exception of a discrepancy in the coefficient $C$.
In Fig~\ref{eos} we plot the equation of state for $^4$He liquid as obtained
by using the values of the fitting parameters reported in 
Table~\ref{fitliq}.
As already known, one notes that both M+AS and M7+AS wavefunctions
give a better equation of state than the shadow wavefunction with
fully optimized Jastrow particle-particle correlations (OJ+AS), although
the OJ+AS wavefunction gives somewhat lower energies.
It has been argued~\cite{macfarland} that possible causes of such
behavior are the incomplete determination of the coefficients in the basis-set
expansion for the OJ+AS wavefunction, or the missing full reoptimization
of the shadow parameters.
Indeed the recent VMC calculations with a fully optimized shadow 
wavefunction~\cite{oswf} confirm this latter possibility.

In Fig.~\ref{geq}, we show the radial distribution function $g(r)$
obtained at the GFMC equilibrium density $\rho \sigma^3=0.365$.
Our maximum of $g(r)$ is obtained at the same position $r_{max}$ as the
GFMC value and it is clear tha there is no shifting of
our curve to larger values of $r$.
Our variational peak $g(r_{max})$ is a little smaller than the
GFMC value.
Fig.~\ref{gfr} shows our results for the radial distribution function 
$g(r)$ determined at the GFMC freezing density $\rho \sigma^3=0.438$.
Statistical errors in the GFMC g(r) near the maximum $g(r_{max})$ are
large, so a detailed comparison with GFMC is not possible.
It appears that the position of our maximum of $g(r)$ compares
very well with the GFMC value, but the variational peak $g(r_{max})$ 
is again smaller.
The trend seen at the GFMC equilibrium and freezing densities is
repeated at all other densities.

In Fig.~\ref{seq} we show $S(k)$ at the equilibrium density
$\rho \sigma^3=0.365$. The experimental $S(k)$ shown in this figure
is the result reported by Svensson {\it et al}\,\cite{svensson} obtained
by neutron diffraction at saturated vapor pressure at $T=1.0 K^o$.
The agreement of the variational structure factor with experiment is seen
to be very good for all $k$-s, except for small $k$.
This is to be expected since our M7+AS wavefunction does not
contain the proper long-range correlations necessary for the linear
behavior~\cite{reattoT} of $S(k)$ which is observed in $^4$He.
%

In this work we have demonstrated the utility of our
M7+AS shadow wavefunction for studying the
ground state properties of liquid and solid $^4$He. The use
of an inverse seventh power as the particle-particle correlation
function has significantly improved the ground state energy, 
the equation of state and the radial density distribution. Such
an improvement was obtained with very little additional computational
effort. The wavefunction remained simple, compact and portable.

\newpage

\begin{table}[t]
\caption[]{ Energies in liquid $^4$He at several
            densities including the experimental equilibrium density
            ($\rho \sigma^3=0.365$)
            and the GFMC freezing density ($\rho \sigma^3=0.438$) with
            $\sigma=2.556 \AA$. 
            All simulations use the Aziz HFDHE2 potential and have been 
            performed for systems of $N=108$ particles.
            The energies are given in Kelvin per particle.	
            M+MS refers to a shadow wavefunction~\cite{vitielloprb}
            with McMillan fifth power-law pseudopotential (m=5) for both
            particle-particle and shadow-shadow pseudopotentials.
            M+AS refers to a shadow wavefunction~\cite{macfarland}
            with a rescaled Aziz HFDHE2 shadow-shadow pseudopotential and  
            a McMillan fifth power-law particle-particle 
            pseudopotential (m=5). 
            M7+AS refers to a shadow wavefunction with a rescaled
            Aziz HFDHE2 shadow-shadow pseudopotential and 
            a McMillan seventh power-law particle-particle 
            pseudopotential (m=7) as used in this work.
	    GFMC refers to the Green's Function Monte Carlo 
            calculations~\cite{gfmc}
            with the Mcmillan fifth power-law form for the importance 
            and starting function. }
\begin{center}
\begin{tabular}{|c|c|c|c|} 
\hline  
$\rho \sigma^3$  &Method &Trial function   &Energy ($K^o$)         \\ \hline   
0.328            &VMC    &M+AS             &-6.561 $\pm$ 0.032     \\ \hline
                 &VMC    &M7+AS        &-6.571 $\pm$ 0.015     \\ \hline
                 &GFMC   & $\cdots$           &-7.034 $\pm$ 0.037     \\ \hline
0.365            &VMC    &M+MS             &-6.061 $\pm$ 0.025     \\ \hline
                 &VMC    &M+AS             &-6.599 $\pm$ 0.034     \\ \hline
                 &VMC    &M7+AS        &-6.664 $\pm$ 0.021     \\ \hline
                 &GFMC   & $\cdots$           &-7.120 $\pm$ 0.024     \\ \hline
0.401            &VMC    &M+AS             &-6.398 $\pm$ 0.019     \\ \hline
                 &VMC    &M7+AS        &-6.497 $\pm$ 0.012     \\ \hline
                 &GFMC   & $\cdots$          &-6.894 $\pm$ 0.048     \\ \hline
0.438            &VMC    &M+MS             &-5.360 $\pm$ 0.035     \\ \hline
                 &VMC    &M+AS             &-5.871 $\pm$ 0.016     \\ \hline
                 &VMC    &M7+AS        &-6.067 $\pm$ 0.010     \\ \hline
                 &GFMC   & $\cdots$          &-6.564 $\pm$ 0.058     \\ \hline
\end{tabular}
\end{center}
\label{tabliquid}
\end{table}

\begin{table}[t]
\caption[]{ The variational parameters of the M7+AS shadow wavefunction
            used in the simulation of $^4$He liquid with $N=108$ 
            particles at different densities. }
\begin{center}
\begin{tabular}{|c|c|c|c|c|} 
\hline  
$\rho \sigma^3$  &$b/\sigma$    &$C \sigma^3$     &$\tau (K^{-1})$  &$\alpha$
                                                                  \\  \hline
0.328            &1.02          &5.5           &0.088      &0.915  \\ \hline
0.365            &1.01          &5.5           &0.095      &0.915  \\ \hline
0.401            &1.01          &6.0           &0.105      &0.920  \\ \hline
0.438            &1.01          &5.9           &0.110      &0.910   \\ \hline
\end{tabular}
\end{center}
\label{parliq}
\end{table}

\begin{table}[t]
\caption[]{ Energies in solid $^4$He at several
            densities including the GFMC melting density 
            ($\rho \sigma^3=0.491$) with $\sigma=2.556 \AA$. 
            All simulations use the Aziz HFDHE2 potential and have been 
            performed for systems of $N=108$ particles.
            The energies are given in Kelvin per particle.
            The notation is the same as in Table~\ref{tabliquid}.
            The GFMC result at density
            $\rho \sigma^3=0.550$ was interpolated from the GFMC
            results at $\rho \sigma^3=0.526$ and $\rho \sigma^3=0.560$.
            The M7+AS results represent this work. }
\begin{center}
\begin{tabular}{|c|c|c|c|} 
\hline  
$\rho \sigma^3$  &Method &Trial function   &Energy ($K^o$)         \\ \hline   
0.491            &VMC    &M+MS             &-5.004 $\pm$ 0.055     \\ \hline
                 &VMC    &M+AS             &-5.052 $\pm$ 0.014     \\ \hline
                 &VMC    &M7+AS        &-5.324 $\pm$ 0.010     \\ \hline
                 &GFMC   & $\cdots$        &-5.610 $\pm$ 0.030     \\ \hline
0.550            &VMC    &M+MS             &-3.521 $\pm$ 0.032     \\ \hline
                 &VMC    &M+AS             &-3.639 $\pm$ 0.012     \\ \hline
                 &VMC    &M7+AS        &-3.724 $\pm$ 0.017     \\ \hline
                 &GFMC   & $\cdots$        &-4.197 $\pm$ 0.030     \\ \hline
0.589            &VMC    &M+AS             &-1.947 $\pm$ 0.012     \\ \hline
                 &VMC    &M7+AS        &-2.097 $\pm$ 0.010     \\ \hline
                 &GFMC   & $\cdots$        &-2.680 $\pm$ 0.060     \\ \hline
\end{tabular}
\end{center}
\label{tabsolid}
\end{table}

\begin{table}[t]
\caption[]{ The variational parameters of the M7+AS shadow 
            wavefunction
            used in the simulation of $^4$He solid with $N=108$ 
            particles at different densities. }
\begin{center}
\begin{tabular}{|c|c|c|c|c|} 
\hline  
$\rho \sigma^3$  &$b/\sigma$    &$C \sigma^3$     &$\tau (K^{-1})$  &$\alpha$
                                                                  \\  \hline
0.491            &1.00          &5.7           &0.110      &0.875  \\ \hline
0.550            &1.00          &5.9           &0.100      &0.890  \\ \hline
0.589            &1.00          &6.5           &0.110      &0.900   \\ \hline
\end{tabular}
\end{center}
\label{parsol}
\end{table}

\begin{table}[t]
\caption[]{ Fit parameters of the equation of state for $^4$He in
            the liquid phase. 
            The OJ+AS shadow wavefunction~\cite{macfarland}
            incorporates a fully optimized Jastrow particle-particle
            pseudopotential.
            The GFMC result is taken from Kalos at al~\cite{gfmc}.
            The experimental equation of state (Exp) is taken from
            Roach {\it et al}\,\cite{roach}. }
\begin{center}
\begin{tabular}{|c|c|c|c|c|} 
\hline  
&           $E_0 (K^o)$            &$B (K^o)$           &$C (K^o)$     
            &$\rho_0 \sigma^3$         \\  \hline
M+AS        &-6.610 $\pm$ 0.036     &10.3 $\pm$ 5.5     & 11.3 $\pm$ 18.5
            &0.3535 $\pm$ 0.0043       \\ \hline
OJ+AS       &-6.796 $\pm$ 0.025     &14.10 $\pm$ 4.18   & -18.7 $\pm$ 18.1
            &0.3567 $\pm$ 0.0032       \\ \hline
M7+AS   &-6.662 $\pm$ 0.020     &14.08 $\pm$ 1.36   & 6.72 $\pm$ 8.02
            &0.361  $\pm$ 0.001       \\ \hline
GFMC        &-7.110 $\pm$ 0.023     &10.08 $\pm$ 3.20   & 12.59 $\pm$ 8.50
            &0.3600 $\pm$ 0.0049       \\ \hline
Exp         &-7.14                  &13.65              & 7.67
            &0.365                      \\ \hline
\end{tabular}
\end{center}
\label{fitliq}
\end{table}

\begin{table}[t]
\caption[]{ Fit parameters of the equation of state for $^4$He in
            the solid phase. 
            The notation is the same as in Table~\ref{fitliq}.
            The fitting curve is 
            $E=E_0+B [(\rho-\rho_s)/\rho_s]^2+C [(\rho-\rho_s)/\rho_s]^3$,
            where $\rho_s \sigma^3=0.4486$ is taken from the GFMC
            result. }
\begin{center}
\begin{tabular}{|c|c|c|c|c|} 
\hline  
&           $E_0 (K^o)$            &$B (K^o)$           &$C (K^o)$     
            &$\rho_s \sigma^3$         \\  \hline
M+AS        &-5.340 $\pm$ 0.021    &31.00 $\pm$ 1.50    & 9.92 $\pm$ 4.34
            &0.4486 $\pm$ 0.0097       \\ \hline
OJ+AS       &-5.81  $\pm$ 0.02     &47.7 $\pm$ 1.4      & -33.89 $\pm$ 4.14
            &0.4486 $\pm$ 0.0097       \\ \hline
M7+AS   &-5.693 $\pm$ 0.013    &32.92 $\pm$ 1.00    & -13.92 $\pm$ 1.45
            &0.4486 $\pm$ 0.0097       \\ \hline
GFMC        &-5.899 $\pm$ 0.121    &31.95 $\pm$ 5.26    & 3.395 $\pm$ 80.0
            &0.4486 $\pm$ 0.0097       \\ \hline
\end{tabular}
\end{center}
\label{fitsol}
\end{table}

\newpage

\begin{figure}[ht]
\caption[] {Equation of state of $^4$He liquid.
            The lines are polynomial cubic fits.
            The solid line is the experimental results~\cite{roach}.
            The dotted line denotes the GFMC result~\cite{gfmc}.
            The dashed line denotes OJ+AS~\cite{macfarland}.
            The solid line with opaque circles is the M7+AS result.
            The solid line with crosses is the 
            M+AS result~\cite{macfarland}. }
\label{eos}
\end{figure}

\begin{figure}[ht]
\caption[] {Radial distribution function $g(r)$ at the
            GFMC equilibrium density $\rho \sigma^3=0.365$ after a VMC
            simulation with N=108 particles.
            The solid line denotes $M7+AS$.
            Filled circles are the GFMC results of 
            Kalos {\it et al}\,\cite{gfmc}.}
\label{geq}
\end{figure}

\begin{figure}[ht]
\caption[] {Radial distribution function $g(r)$ at the
            GFMC freezing density $\rho \sigma^3=0.438$ after a VMC
            simulation with N=108 particles.
            The solid line denotes $M7+AS$.
            Filled circles are the GFMC results of 
            Kalos {\it et al}\,\cite{gfmc}.}
\label{gfr}
\end{figure}

\begin{figure}[ht]
\caption[]{Static structure factor $S(k)$ of liquid $^4$He 
           at the GFMC equilibrium density $\rho \sigma^3=0.365$.
           The filled circles show our results obtained
           from the formula
           $S(k)=\frac{1}{N} \langle \rho_{- {\bf k}} \rho_{ {\bf k}} \rangle$
           at a discrete set of $k$-points.
           The solid line denotes the experimental
           results reported by Svensson and co-workers~\cite{svensson}
           obtained at saturated vapor pressure by means of neutron
           diffraction at temperature $T=1.0 K^o$.}
\label{seq}
\end{figure}

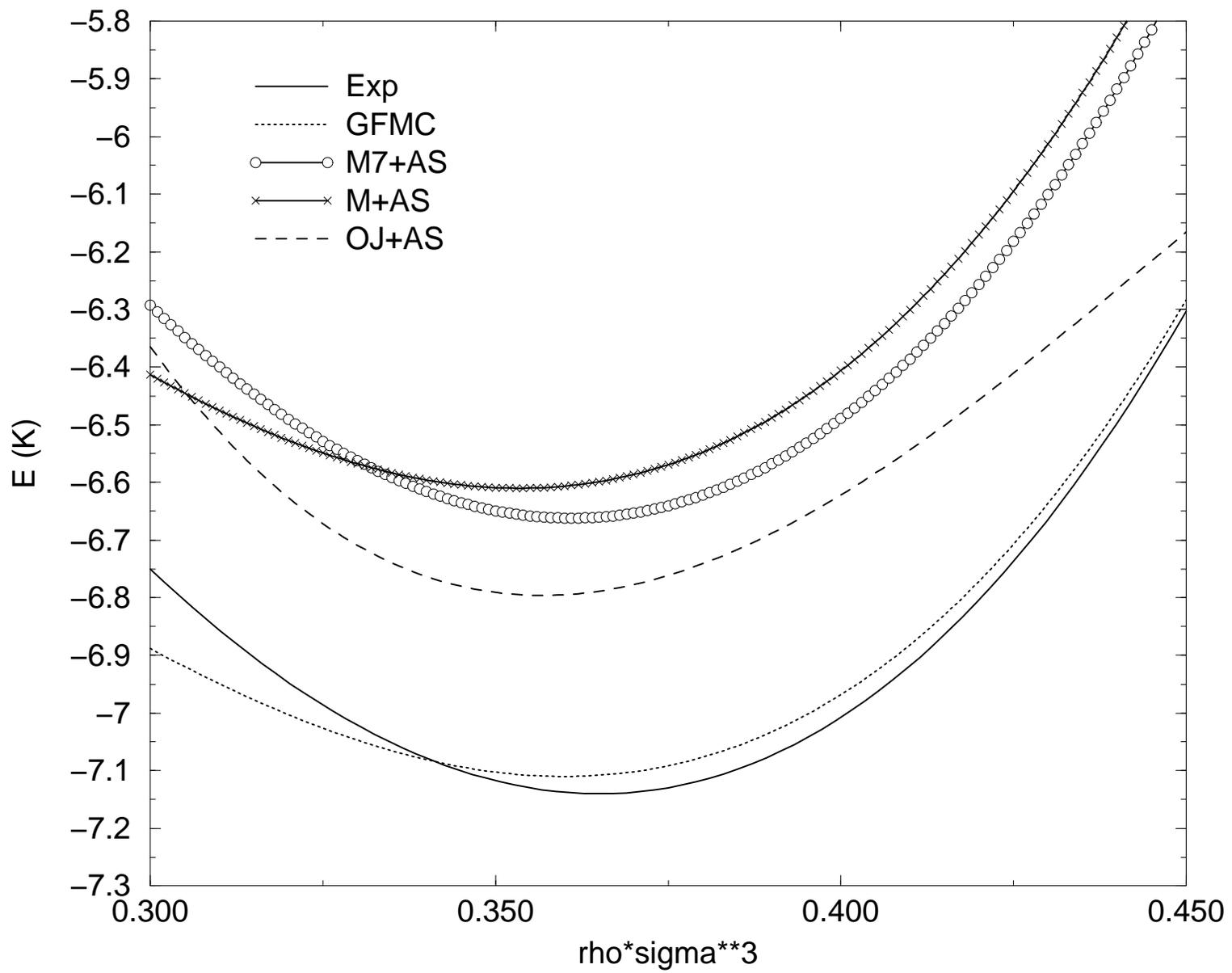
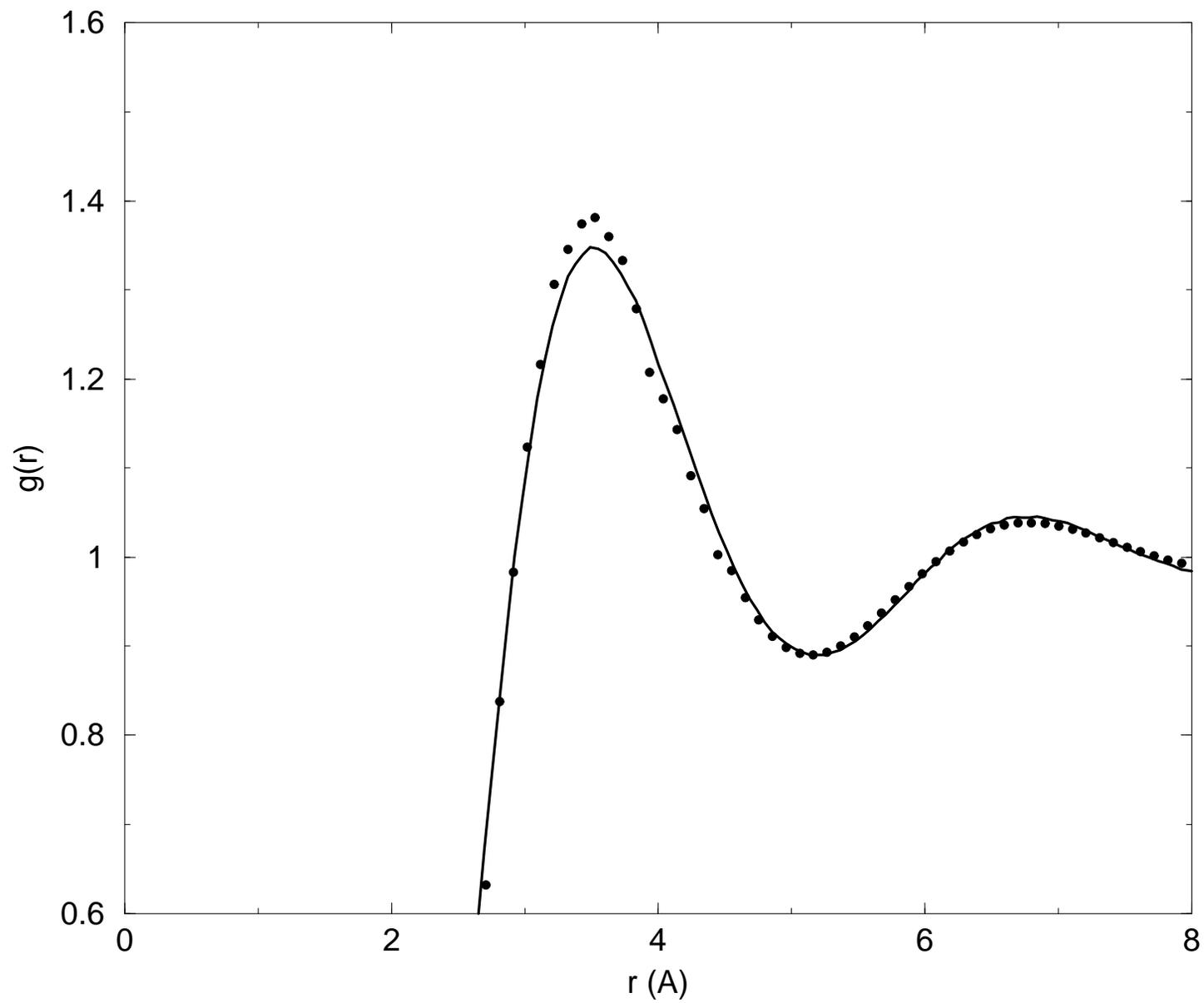
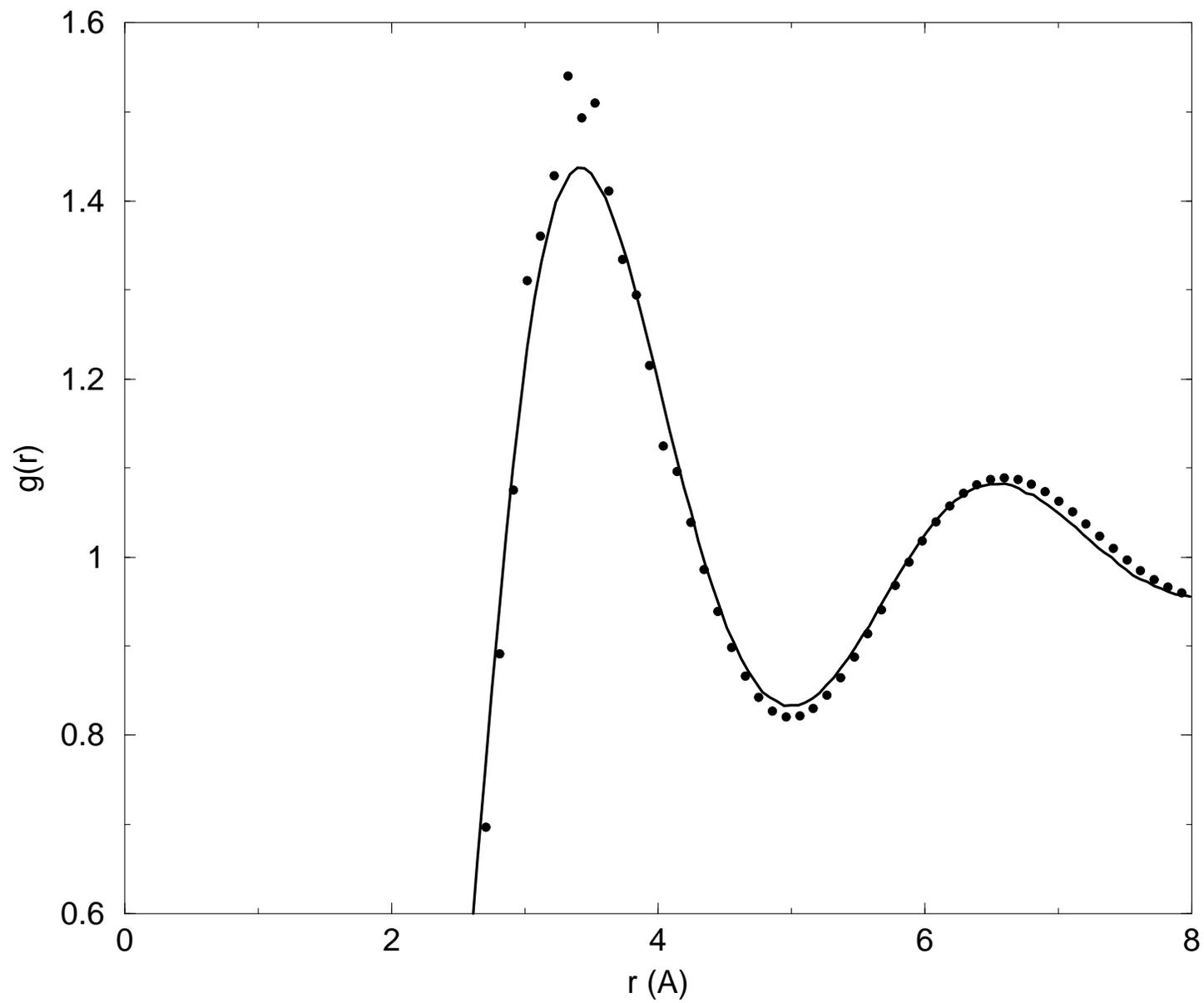
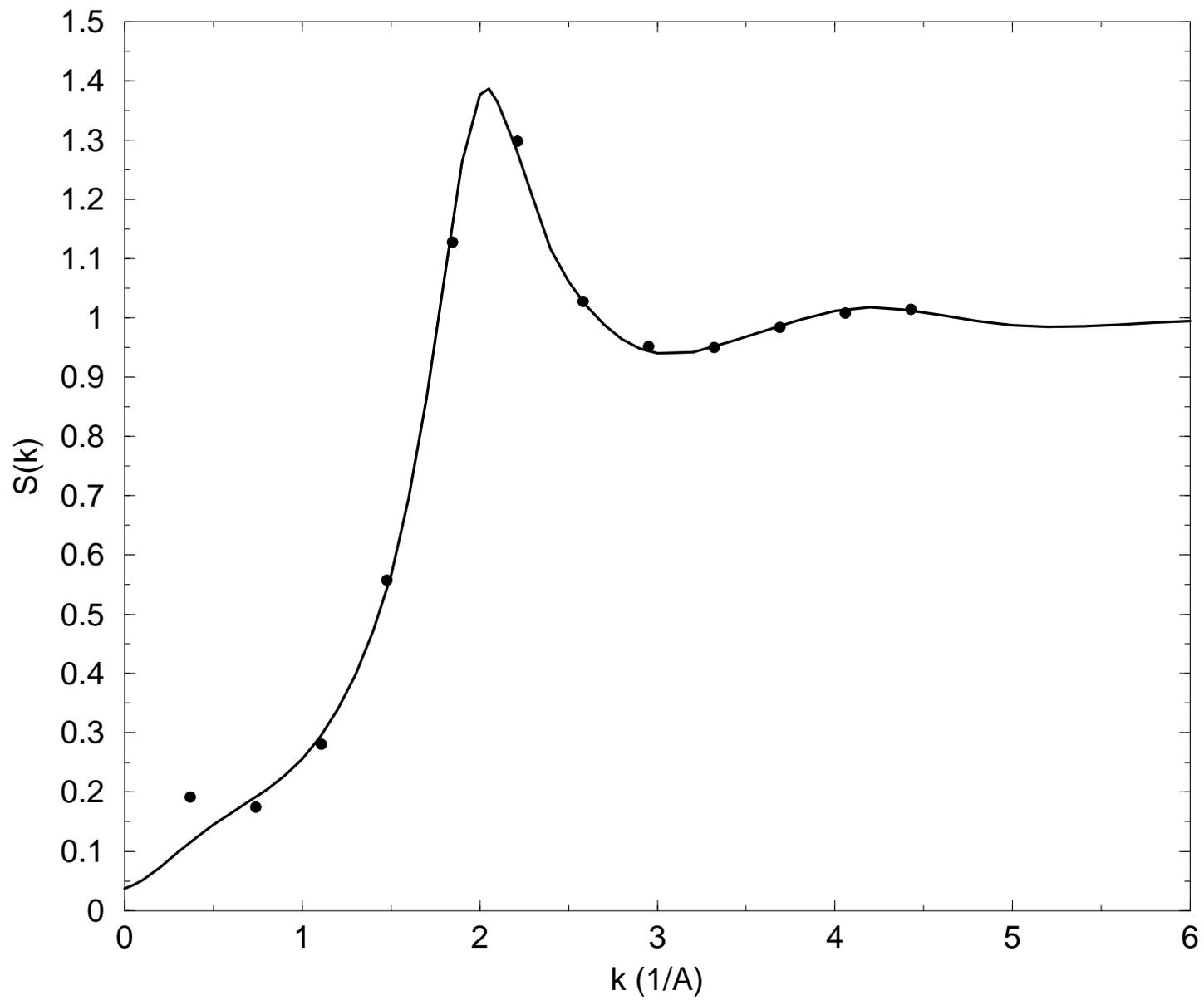
\end{document}